\documentclass[aps,prb,twocolumn,showpacs]{revtex4}

\usepackage{graphicx}
\usepackage{amsmath}
\usepackage{amssymb}

\begin{document}

\title{Renormalization of dispersion in electron-doped bilayer cuprate superconductors}

\author{Shuning Tan$^{1}$\footnote{E-mail: sntan@ysu.edu.cn}, Yiqun Liu$^{2}$, Yingping 
Mou$^{3}$, Huaiming Guo$^{4}$, and Shiping Feng$^{5}$\footnote{E-mail: spfeng@bnu.edu.cn}}

\affiliation{$^{1}$Key Laboratory for Microstructural Material Physics of Hebei Province, 
School of Science, Yanshan University, Qinhuangdao 066004, China}

\affiliation{$^{2}$School of Physics, Nanjing University, Nanjing 210093, China}

\affiliation{$^{1}$Beijing Computational Science Research Center, Beijing 100193, China}

\affiliation{$^{4}$School of Physics, Beihang University, Beijing 100191, China}

\affiliation{$^{5}$Department of Physics, Beijing Normal University, Beijing 100875, China}

\begin{abstract}
The renormalization of the electrons in cuprate superconductors is characterized by the kink in the
quasiparticle dispersion. Here the bilayer coupling effect on the quasiparticle dispersion kink in
the electron-doped bilayer cuprate superconductors is studied based on the kinetic-energy-driven
superconductivity. It is shown that the kink in the quasiparticle dispersion is present all around
the electron Fermi surface, as the quasiparticle dispersion kink in the single-layer case. However,
in comparison with the corresponding single-layer case, the kink effect in the quasiparticle
dispersion at around the antinodal region becomes the most pronounced, indicating that the kink
effect in the quasiparticle dispersion at around the antinodal region is enhanced by the bilayer
coupling.
\end{abstract}

\pacs{74.25.Jb, 74.25.Dw, 74.20.Mn, 74.72.Ek}

\maketitle

The parent compounds of cuprate superconductors are structurally characterized by the stacking
CuO$_{2}$ layers \cite{Bednorz86,Tokura89}, and then superconductivity emerges when charge
carriers, holes or electrons, are doped into these CuO$_{2}$ layers
\cite{Damascelli03,Campuzano04,Fink07,Armitage10}. In both the electron- and hole-doped cuprate
superconductors, the renormalization of the electrons to form the quasiparticles are due to the
strong electron's coupling to various bosonic excitations \cite{Carbotte11,Bok16}, and then the
unconventional properties, including the exceptionally high superconducting (SC) transition
temperature $T_{\rm c}$, have often been attributed to particular characteristics of the
quasiparticle excitations \cite{Damascelli03,Campuzano04,Fink07,Armitage10,Carbotte11,Bok16,Yin21}.

The renormalization of the electrons is characterized by the kink in the quasiparticle dispersion
\cite{Damascelli03,Campuzano04,Fink07,Armitage10}. In the hole-doped case, the kink in the
quasiparticle dispersion was firstly observed in the angle-resolved photoemission spectroscopy
(ARPES) measurements on the hole-doped bilayer cuprate superconductors
\cite{Bogdanov00,Kaminski01,Johnson01,Sato03,Kordyuk06}. Later, this quasiparticle dispersion kink
is found to be present in all families of the hole-doped cuprate superconductors with one or more
CuO$_{2}$ layers per unit cell \cite{Zhou03,Yoshida07,Lee09,Chen09}. In particular, these
experimental observations also show that the quasiparticle dispersion kink is particularly obvious
in these hole-doped cuprate superconductors with two or more CuO$_{2}$ layers per unit cell
\cite{Bogdanov00,Kaminski01,Johnson01,Sato03,Kordyuk06,Zhou03,Yoshida07,Lee09,Chen09}, reflecting
an experimental fact that the kink effect in the quasiparticle dispersion is enhanced by the strong
coupling between the CuO$_{2}$ layers in a unit cell. On the electron-doped side, both the doped
electrons and the proper annealing process in a low-oxygen environment are required to induce the
intrinsic aspects of the unconventional properties \cite{Armitage10,Adachi13,Adachi17}. In this case,
although the kink in the quasiparticle dispersion associated with the improper annealing condition
has been observed early \cite{Santander-Syro09,Schmitt08,Park08a,Armitage03}, the intrinsic kink
effect associated with the proper annealing condition was observed recently \cite{Horio20b}. Moreover,
the evolution of the electronic structure with the electron doping in the electron-doped bilayer
cuprate superconductors has been investigated very recently in terms of the ARPES measurements
\cite{Hu21}, where the entire momentum-resolved energy spectrum in the lightly electron doped case
has been directly visualized. This electron-doped bilayer cuprate superconductor is an ideal system
to tackle the coupling effect between the CuO$_{2}$ layers within a unit cell on the renormalization
of the dispersion \cite{Hu21}, however, the experimental data of the quasiparticle dispersion kink in
the electron-doped bilayer cuprate superconductors is still lacking to date. In other words, it is
still unclear whether the kink effect in the quasiparticle dispersion of the electron-doped cuprate
superconductors is enhanced by the coupling between the CuO$_{2}$ layers within a unit cell or not?

In our recent work \cite{Tan21}, the quasiparticle dispersion kink in the electron-doped single-layer
cuprate superconductors has been studied based on the kinetic-energy-driven superconductivity, where
we have shown that although the optimized $T_{\rm c}$ in the electron-doped single-layer cuprate
superconductors is much smaller than that in the hole-doped counterparts, the electron- and hole-doped
single-layer cuprate superconductors rather resemble each other in the doping range of the SC dome,
indicating an absence of the disparity between the phase diagrams of the electron- and hole-doped
cuprate superconductors. Moreover, the quasiparticle dispersion is affected by the spin excitation,
and then the kink in the quasiparticle dispersion is always accompanied by the corresponding inflection
point in the total self-energy \cite{Tan21}. In this paper, we study the effect of the coupling between
the CuO$_{2}$ layers within a unit cell on the quasiparticle dispersion kink in the electron-doped
bilayer superconductors along with this line, and show that the quasiparticle dispersion kink is present
all around the electron Fermi surface (EFS), as the corresponding quasiparticle dispersion kink in the
single-layer case \cite{Tan21}. However, in comparison with the corresponding single-layer case
\cite{Tan21}, the kink effect in the quasiparticle dispersion at around the antinodal region becomes the
most pronounced, which therefore indicates that the kink effect in the quasiparticle dispersion at
around the antinodal region is enhanced by the bilayer coupling.

It is commonly accepted that the essential physics of the doped CuO$_{2}$ layer can be described by the
$t$-$J$ model on a square lattice \cite{Anderson87}. However, for the discussions of the bilayer
coupling effect on the quasiparticle dispersion kink in the electron-doped bilayer cuprate
superconductors, the single-layer $t$-$J$ model can be extended by the consideration of the bilayer
coupling as \cite{Liu20},
\begin{eqnarray}\label{bilayer-tJ-model}
H&=&-t\sum_{l\hat{\eta}a\sigma}C^{\dagger}_{la\sigma}C_{l+\hat{\eta}a\sigma}
+t'\sum_{l\hat{\tau}a\sigma}C^{\dagger}_{la\sigma}C_{l+\hat{\tau}a\sigma}\nonumber\\
&-&\sum_{la\neq b\sigma}t_{\perp}(l)C^{\dagger}_{la\sigma}C_{lb\sigma}+\mu\sum_{la\sigma}
C^{\dagger}_{la\sigma}C_{la\sigma} \nonumber\\
&+&J\sum_{l\hat{\eta}a}{\bf S}_{la}\cdot {\bf S}_{l+\hat{\eta}a}+J_{\perp}
\sum_{la\neq b}{\bf S}_{la}\cdot {\bf S}_{lb},~~~~
\end{eqnarray}
where $a (b)=1,2$, is the CuO$_{2}$ layer index, the hopping integrals $t<0$ and $t'<0$ in the
electron-doped side, $\hat{\eta}=\pm\hat{x},\pm\hat{y}$ represents the nearest-neighbor (NN) sites of
the site $l$, $\hat{\tau}=\pm\hat{x}\pm\hat{y}$ represents the next NN sites of the site $l$,
$C^{\dagger}_{la\sigma}$ ($C_{la\sigma}$) is the electron creation (annihilation) operator,
${\bf S}_{l}$ is the spin operator with its components $S_{l}^{x}$, $S_{l}^{y}$, and $S_{l}^{z}$, and
$\mu$ is the chemical potential, while the momentum dependence of the bilayer hopping
$t_{\perp}({\bf k})$ is given by \cite{Massida88,Chakarvarty93,Andersen94,Andersen95,Liechtenstein96},
\begin{eqnarray}\label{interlayer-hopping}
t_{\perp}({\bf k})={t_{\perp}\over 4}(\cos k_{x} -\cos k_{y})^{2},
\end{eqnarray}
which therefore leads to that the bilayer magnetic exchange $J_{\perp}=(t_{\perp}/t)^{2}J$.

This bilayer $t$-$J$ model (\ref{bilayer-tJ-model}) is supplemented by the local constraint
$\sum_{\sigma}C^{\dagger}_{la\sigma}C_{la\sigma}\geq 1$ to remove zero electron occupancy, which is
different from the hole-doped case \cite{Liu20}, where the local constraint is subjected to remove
double electron occupancy, and can be treated properly in terms of the charge-spin separation
fermion-spin transformation \cite{Feng9404,Feng15}. However, for the application of this fermion-spin
transformation to treat the local constraint in the electron-doped side, we can work in the hole
representation via a particle-hole transformation $C_{la\sigma}\rightarrow f^{\dagger}_{la-\sigma}$.
In this hole representation, the $t$-$J$ model (\ref{bilayer-tJ-model}) can be rewritten as
\cite{Tan21},
\begin{eqnarray}\label{hole-bilayer-tJ-model}
H&=&t\sum_{l\hat{\eta}a\sigma}f^{\dagger}_{la\sigma}f_{l+\hat{\eta}a\sigma}
-t'\sum_{l\hat{\tau}a\sigma}f^{\dagger}_{la\sigma}f_{l+\hat{\tau}a\sigma}\nonumber\\
&+&\sum_{la\neq b\sigma}t_{\perp}(l)f^{\dagger}_{la\sigma}f_{lb\sigma}-\mu\sum_{la\sigma}
f^{\dagger}_{la\sigma}f_{la\sigma} \nonumber\\
&+&J\sum_{l\hat{\eta}a}{\bf S}_{la}\cdot {\bf S}_{l+\hat{\eta}a}+J_{\perp}
\sum_{la\neq b}{\bf S}_{la}\cdot {\bf S}_{lb},~~~~
\end{eqnarray}
and then the local constraint of no zero electron occupancy in the electron representation
$\sum_{\sigma}C^{\dagger}_{la\sigma}C_{la\sigma}\geq 1$ is replaced by the local constraint of no
double hole occupancy in the hole representation $\sum_{\sigma}f^{\dagger}_{la\sigma}f_{la\sigma}\leq 1$,
where $f^{\dagger}_{la\sigma}$ ($f_{la\sigma}$) is the hole creation (annihilation) operator. According
to the fermion-spin transformation \cite{Feng9404,Feng15}, the constrained hole operators $f_{la\uparrow}$
and $f_{la\downarrow}$ now can be decoupled as: $f_{la\uparrow}=a^{\dagger}_{la\uparrow}S^{-}_{la}$ and
$f_{la\downarrow}=a^{\dagger}_{la\downarrow}S^{+}_{la}$, where the spinful fermion operator
$a_{la\sigma}=e^{-i\Phi_{l\sigma}}a_{la}$ keeps track of the charge degree of freedom of the constrained
hole together with some effects of spin configuration rearrangements due to the presence of the doped
charge carrier itself, while the localized spin operator $S_{l}$ keeps track of the spin degree of
freedom of the constrained hole, and then the local constraint without double hole occupancy is satisfied
in analytical calculations.

Within the bilayer $t$-$J$ model in the fermion-spin representation, the renormalization of the dispersion
in the hole-doped bilayer cuprate superconductors \cite{Liu20} has been investigated based on the
kinetic-energy-driven superconductivity \cite{Feng15,Feng0306,Feng12,Feng15a}, where the kink effect in
the quasiparticle dispersion is enhanced by the strong bilayer coupling. Following these previous
discussions \cite{Liu20}, the hole normal and anomalous Green's functions of the bilayer $t$-$J$ model
(\ref{hole-bilayer-tJ-model}) in the bonding-antibonding representation can be obtained explicitly as,
\begin{subequations}\label{HGFS}
\begin{eqnarray}
G^{\rm (f)}_{\nu}({\bf k},\omega)&=&{1\over\omega-\varepsilon^{\rm (f)}_{{\nu}{\bf k}}
-\Sigma_{{\nu}{\rm tot}}^{\rm (f)}({\bf k},\omega)},\label{NHGF}\\
\Im^{{\rm (f)}\dagger}_{\nu}({\bf k},\omega)&=&{L_{\nu}^{(\rm f)}({\bf k},\omega)\over
\omega-\varepsilon^{\rm (f)}_{\nu{\bf k}}-\Sigma_{\nu{\rm tot}}^{\rm (f)}({\bf k},\omega)},
\label{ANHGF}
\end{eqnarray}
\end{subequations}
where $\nu=1,2$ with $\nu=1$ ($\nu=2$) that represents the corresponding bonding (antibonding) component,
the bare hole dispersion
$\varepsilon^{\rm (f)}_{\nu{\bf k}}=\varepsilon^{\rm (f)}_{\bf k}+(-1)^{\nu}t_{\perp}({\bf k})$, with
$\varepsilon^{\rm (f)}_{\bf k}=-4t\gamma_{\bf k}+4t'\gamma'_{\bf k}+\mu$,
$\gamma_{\bf k}=(\cos{{\rm k}_x}+\cos{{\rm k}_y})/2$, and $\gamma'_{\bf k}=\cos{{\rm k}_x}\cos{{\rm k}_y}$,
while the hole total self-energy
$\Sigma^{\rm (f)}_{\nu{\rm tot}}({\bf k},\omega)$ is a specific combination of the hole normal self-energy
$\Sigma^{\rm (f)}_{\nu{\rm ph}}({\bf k},\omega)$ in the particle-hole channel and the hole anomalous
self-energy $\Sigma^{\rm (f)}_{\nu{\rm pp}}({\bf k},\omega)$ in the particle-particle channel as,
\begin{eqnarray}
\Sigma^{\rm (f)}_{\nu{\rm tot}}({\bf k},\omega)=\Sigma^{\rm (f)}_{\nu{\rm ph}}({\bf k},\omega)
+{|\Sigma^{\rm (f)}_{\nu{\rm pp}}({\bf k},\omega)|^{2}\over \omega
+\varepsilon^{\rm (f)}_{\nu{\bf k}}+\Sigma^{\rm (f)}_{\nu{\rm ph}}({\bf k},-\omega)},
\label{TOT-HSE}
\end{eqnarray}
and the function $L_{\nu}^{(\rm f)}({\bf k},\omega)$ is given by,
\begin{eqnarray}
L_{\nu}^{(\rm f)}({\bf k},\omega)=-{\Sigma^{\rm (f)}_{\nu{\rm pp}}({\bf k},\omega)\over \omega
+\varepsilon^{\rm (f)}_{\nu{\bf k}}+\Sigma^{\rm (f)}_{\nu{\rm ph}}({\bf k},-\omega)},
\end{eqnarray}
where hole normal self-energy $\Sigma^{\rm (f)}_{\nu{\rm ph}}({\bf k},\omega)$ and hole anomalous
self-energy $\Sigma^{\rm (f)}_{\nu{\rm pp}}({\bf k},\omega)$ have been given explicitly in Ref.
\onlinecite{Liu20} except for $t<0$ and $t'<0$ in the electron-doped side.

Our main goal is to derive explicitly the electron normal and anomalous Green's functions
$G_{\nu}({\bf k},\omega)$ and $\Im^{\dagger}_{\nu}({\bf k},\omega)$ of the $t$-$J$ model
(\ref{bilayer-tJ-model}), which are directly associated with the hole normal and anomalous Green's
functions $G^{\rm (f)}_{\nu}({\bf k},\omega)$ and $\Im^{{\rm (f)}\dagger}_{\nu}({\bf k},\omega)$ in
Eq. (\ref{HGFS}) via the particle-hole transformation $C_{la\sigma}\rightarrow f^{\dagger}_{la-\sigma}$
as $G(l-l',t-t')=\langle\langle C_{l\sigma}(t);C^{\dagger}_{l'\sigma}(t')\rangle\rangle=\langle\langle f^{\dagger}_{l\sigma}(t);f_{l'\sigma}(t')\rangle\rangle=-G^{\rm (f)}(l'-l, t'-t)$ and
$\Im(l-l',t-t')=\langle\langle C_{l\downarrow}(t); C_{l'\uparrow}(t')\rangle\rangle=\langle\langle f^{\dagger}_{l\uparrow}(t);f^{\dagger}_{l'\downarrow}(t')\rangle\rangle=\Im^{{\rm (f)}\dagger}(l-l',t-t')$.
With the helps of the above hole normal and anomalous Green's functions (\ref{HGFS}), the electron normal
and anomalous Green's functions in the bonding-antibonding representation can be derived as \cite{Tan21}
$G_{\nu}({\bf k},\omega) =-G^{\rm (f)}_{\nu}({\bf k},-\omega)$ and
$\Im_{\nu}({\bf k},\omega)=\Im^{{\rm (f)}\dagger}_{\nu}({\bf k}, \omega)$, respectively, with the bare
electron dispersion $\varepsilon^{(\nu)}_{\bf k}$, the electron normal self-energy
$\Sigma^{(\nu)}_{\rm ph}({\bf k},\omega)$, and the electron anomalous self-energy
$\Sigma^{(\nu)}_{\rm pp}({\bf k},\omega)$ that can be obtained as
$\varepsilon^{(\nu)}_{\bf k}=-\varepsilon^{({\rm f})}_{\nu{\bf k}}$,
$\Sigma^{(\nu)}_{\rm ph}({\bf k},\omega)=-\Sigma^{({\rm f})}_{\nu{\rm ph}}({\bf k},-\omega)$,
and $\Sigma^{(\nu)}_{\rm pp}({\bf k},\omega)=\Sigma^{({\rm f})}_{\nu{\rm pp}}({\bf k},\omega)$,
respectively. In this case, the electron spectral function
$A_{\nu}({\bf k},\omega)=-2{\rm Im}G_{\nu}({\bf k},\omega)$ in the bonding-antibonding representation now
can be obtained explicitly as,
\begin{eqnarray}\label{ESF}
A_{\nu}({\bf k},\omega)={-2{\rm Im}\Sigma^{(\nu)}_{\rm tot}({\bf k},\omega)\over
[\omega-\varepsilon^{(\nu)}_{\bf k}-{\rm Re}\Sigma^{(\nu)}_{\rm tot}({\bf k},\omega)]^{2}
+[{\rm Im}\Sigma^{(\nu)}_{\rm tot}({\bf k},\omega)]^{2}},~
\end{eqnarray}
with the electron total self-energy $\Sigma^{(\nu)}_{\rm tot}({\bf k},\omega)$,
\begin{eqnarray}
\Sigma^{(\nu)}_{\rm tot}({\bf k},\omega)=\Sigma^{(\nu)}_{\rm ph}({\bf k},\omega)
+{|\Sigma^{(\nu)}_{\rm pp}({\bf k},\omega)|^{2}\over \omega
+\varepsilon^{(\nu)}_{\bf k}+\Sigma^{(\nu)}_{\rm ph}({\bf k},-\omega)},
\label{TOT-ESE}
\end{eqnarray}
where ${\rm Re}\Sigma^{(\nu)}_{\rm tot}({\bf k},\omega)$ and
${\rm Im}\Sigma^{(\nu)}_{\rm tot}({\bf k},\omega)$ are the real and imaginary parts of the electron total
self-energy, respectively. As we have shown in the previous studies \cite{Tan21} that in the numerical
calculation at a finite lattice, the sharp peak visible for temperature $T\rightarrow 0$ in the electron
normal (anomalous) self-energy is actually a $\delta$-functions, broadened by a small damping. The
calculation in this paper for the electron normal (anomalous) self-energy is performed numerically on a
$120\times 120$ lattice in momentum space, with the infinitesimal $i0_{+}\rightarrow i\Gamma$ replaced by
a small damping $\Gamma=0.05J$. In the following discussions, we fix the parameters $t/J=-2.5$,
$t'/t=0.22$, and $t_{\perp}/t=0.1$ in the bilayer $t$-$J$ model (\ref{bilayer-tJ-model}), and then all
the energy scales are in the units of $J=1$. However, to compare with the single-layer case, we use
$J=100$ meV. All these values of the parameters are the typical values of the electron-doped cuprate
superconductors \cite{Armitage10,Hu21}.

\begin{figure}[h!]
\centering
\includegraphics[scale=1.2]{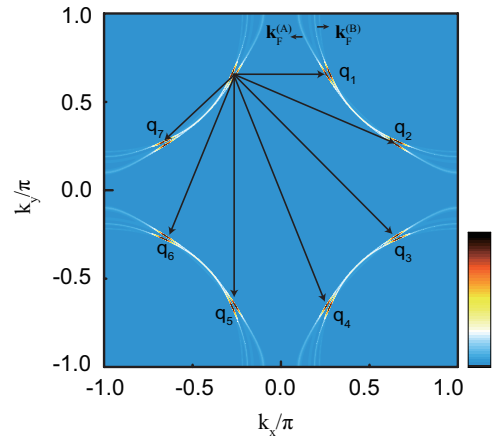}
\caption{(Color online) The intensity map of the bonding and antibonding components of the electron
spectral function $A_{1}({\bf k},0)$ and $A_{2}({\bf k},0)$ at zero binding-energy in $\delta=0.15$
with $T=0.002J$ for $t/J=-2.5$, $t'/t=0.22$, and $t_{\perp}/t=0.1$. \label{spectral-maps}}
\end{figure}

In cuprate superconductors, the topology of EFS plays an essential role in the understanding of the
unconventional properties, since everything happens at around EFS. In particular, the strong coupling
between the electrons and a strongly dispersive spin excitation in cuprate superconductors leads to a
strong redistribution of the spectral weights on EFS \cite{Tan21,Liu20}, and then a bewildering
variety of electronically ordered states is driven by this EFS instability
\cite{Neto15,Neto16,Horio16,Mou17}. However, as shown in Eq. (\ref{ESF}), the electron spectral function
in the electron-doped bilayer cuprate superconductors has been separated into its bonding and antibonding
components due to the presence of the bilayer coupling, which leads to two EFS contours
${\bf k}^{\rm (B)}_{\rm F}$ and ${\bf k}^{\rm (A)}_{\rm F}$ deriving directly from the bonding and
antibonding layers. For a convenience in the following discussions of the bilayer coupling effect on the
quasiparticle dispersion kink, we firstly plot the EFS map from both the bonding component
$A_{1}({\bf k},0)$ and the antibonding component $A_{2}({\bf k},0)$ of the electron spectral function
(\ref{ESF}) at zero binding-energy for $\delta=0.15$ with $T=0.002J$ in Fig. \ref{spectral-maps}, where
in corresponding to the momentum dependence of the interlayer hopping in Eq. (\ref{interlayer-hopping}),
the maximal distance between the bonding and antibonding EFS contours ${\bf k}^{\rm (B)}_{\rm F}$ and
${\bf k}^{\rm (A)}_{\rm F}$ appears at around the antinodal region, then it gradually reduces when the
momentum moves away from the antinodal region, and it eventually disappears at around the nodal region,
indicating that the bilayer coupling with the high impacts on the electronic structure mainly occurs at
around antinodal region. Apart from this intrinsic feature, another intrinsic feature is the
redistribution of the spectral weights in the bonding and antibonding EFS contours, where the spectral
weight on ${\bf k}^{\rm (B)}_{\rm F}$ (${\bf k}^{\rm (A)}_{\rm F}$) at around the bonding (antibonding)
antinodal region is reduced greatly. As a natural consequence, the bonding (antibonding) EFS contour is
broken up into the disconnected bonding (antibonding) Fermi arcs located at around the bonding
(antibonding) nodal region. However, the renormalization from the quasiparticle scattering further
reduces the most part of the spectral weight on the bonding (antibonding) Fermi arcs to the tips of the
bonding (antibonding) Fermi arcs \cite{Horio16,Mou17}, and then these tips of the Fermi arcs connected
by the scattering wave vectors ${\bf q}_{i}$ shown in Fig. \ref{spectral-maps} construct an octet
scattering model. In this case, a bewildering variety of electronic orders described by the quasiparticle
scattering processes with the scattering wave vectors ${\bf q}_{i}$ therefore is driven by this EFS
instability \cite{Neto15,Neto16,Horio16,Mou17}. These intrinsic features of the EFS reconstruction
resemble these obtained for the corresponding hole-doped case \cite{Liu20}.

\begin{figure}[h!]
\centering
\includegraphics[scale=0.83]{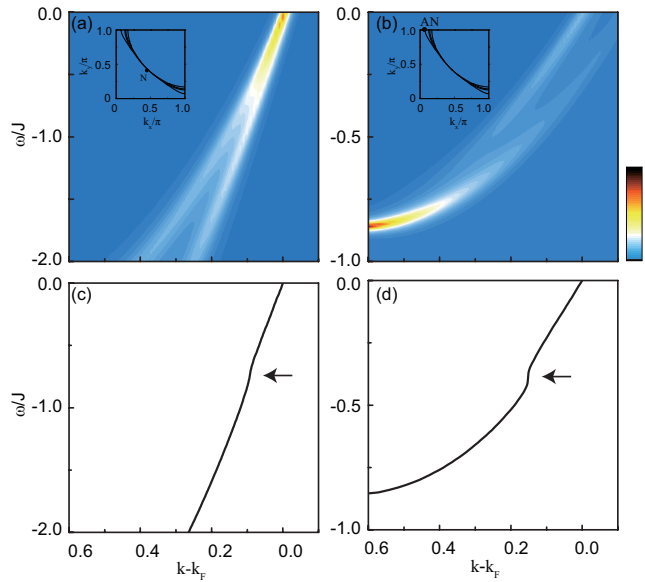}
\caption{(Color online) Upper panel: the intensity maps of the antibonding component of the electron
spectral function as a function of binding-energy along (a) the nodal cut and (b) the antinodal cut
at $\delta=0.15$ with $T=0.002J$ for $t/J=-2.5$, $t'/t=0.22$, and $t_{\perp}/t=0.1$. Lower panel: the
antibonding quasiparticle dispersions along (c) the nodal cut and (b) the antinodal cut extracted
from the positions of the lowest-energy quasiparticle excitation peaks in (a) and (b), respectively.
The arrow indicates the kink position. \label{kink-maps}}
\end{figure}

We are now ready to discuss the bilayer coupling effect on the quasiparticle dispersion kink in the
electron-doped bilayer cuprate superconductors. In the upper panel of Fig. \ref{kink-maps}, we plot the
intensity map of the antibonding component $A_{2}({\bf k},\omega)$ of the electron spectral function as a
function of binding-energy along (a) the nodal cut and (b) the antinodal cut at $\delta=0.15$ with
$T=0.002J$, while the corresponding antibonding quasiparticle dispersions along (c) the nodal cut and (d)
the antinodal cut extracted from the positions of the lowest-energy quasiparticle excitation peaks in (a)
and (b), respectively, are shown in the lower panel. It is shown clearly that although the kink in the
quasiparticle dispersion is present all around the antibonding EFS, the quasiparticle dispersive behavior
at around the nodal region is much different from that at around the antinodal region. The results in
Fig. \ref{kink-maps}a and Fig. \ref{kink-maps}c show that the quasiparticle dispersion along the nodal
cut at both the low binding-energy and high binding-energy ranges exhibits a linear behavior, but with
different slopes, and then these two ranges with different slopes are separated by a weak kink. In
particular, for the low binding-energy less than the kink energy, the spectrum exhibits sharp peaks with
a weak dispersion, while for the high binding-energy greater than the kink energy, the spectrum exhibits
broad peaks with a stronger dispersion, as the renormalization of the dispersion in the single-layer case
\cite{Tan21}. However, the kink effect in the quasiparticle dispersion of the electron-doped bilayer
cuprate superconductors at around the the antinodal region is more stronger than that in the single-layer
case. This follows a basic fact that at the kink energy, although the quasiparticle dispersion of the
electron-doped single-layer cuprate superconductors at around the antinodal region is broken up into a
fast dispersive high binding-energy part and a slow dispersive low binding-energy part \cite{Tan21},
this breaking effect becomes the most pronounced in the electron-doped bilayer cuprate superconductors
as shown in Fig. \ref{kink-maps}b and Fig. \ref{kink-maps}d, which shows that the kink effect in the
quasiparticle dispersion of the electron-doped bilayer cuprate superconductors at around the antinodal
region is enhanced. Concomitantly, the quasiparticle dispersion kink emerges in the energy
$\omega_{\rm kink}\sim 0.72J=72$ meV at around the nodal region, while the quasiparticle dispersion kink
occurs in the energy $\omega_{\rm kink}\sim 0.39J=39$ meV at around the antinodal region, which therefore
indicates that the characteristic kink energy decreases gradually when the momentum moves from the nodal
region to the antinodal region, in qualitative agreement with these in the single-layer case \cite{Tan21}.
Our present result of the enhancement of the kink effect in the quasiparticle dispersion of the
electron-doped bilayer cuprate superconductors is also qualitatively consistent with that in the
hole-doped bilayer cuprate superconductors \cite{Liu20}, suggesting a common quasiparticle dispersion kink
mechanism for both the electron- and hole-doped cuprate superconductors.

\begin{figure}[h!]
\centering
\includegraphics[scale=0.85]{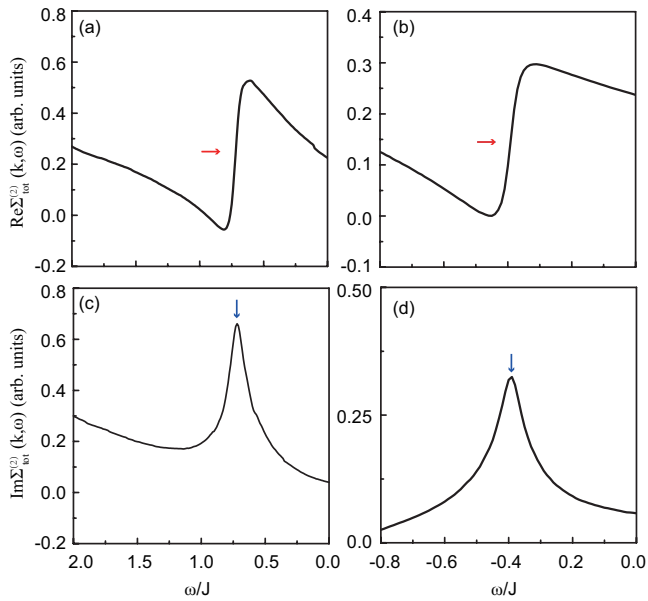}
\caption{(Color online) Upper panel: the real part of the antibonding total self-energy as a function
of binding-energy along (a) the nodal dispersion and (b) the antinodal dispersion at $\delta=0.15$
with $T=0.002J$ for $t/J=-2.5$, $t'/t=0.22$, and $t_{\perp}/t=0.1$. Lower panel: the corresponding
imaginary part of the antibonding total self-energy as a function of binding-energy along (c) the
nodal dispersion and (d) the antinodal dispersion. The red arrow indicates inflection point, while
the blue arrow denotes the peak position. \label{band-structure}}
\end{figure}

The essential physics of the kink effect in the quasiparticle dispersion of the electron-doped bilayer
cuprate superconductors and of its enhancement at around the antinodal region is the same as that in the
corresponding hole-doped case \cite{Liu20}. On the one hand, the emergence of the quasiparticle
dispersion kink in the electron-doped bilayer cuprate superconductors can be also attributed to the
electron self-energy effects arising from the strong electron's coupling to a strongly dispersive spin
excitation, as the emergence of the quasiparticle dispersion kink in the single-layer case \cite{Tan21}.
To see this point more clearly, we plot the real part of the antibonding total self-energy
${\rm Re}\Sigma^{(2)}_{\rm tot}({\bf k},\omega)$ as a function of binding-energy along (a) the nodal
dispersion and (b) the antinodal dispersion as shown in Fig. \ref{kink-maps}c and Fig. \ref{kink-maps}d,
respectively, at $\delta=0.15$ with $T=0.002J$ in the upper panel of Fig. \ref{band-structure}, where the
red arrow indicates the inflection point (then the point of the slope change). In the lower panel, we plot
the corresponding imaginary part of the antibonding total self-energy
${\rm Im}\Sigma^{(2)}_{\rm tot}({\bf k},\omega)$ (then the antibonding quasiparticle scattering rate) as a
function of binding-energy along (c) the nodal dispersion and (d) the antinodal dispersion, where the blue
arrow denotes the peak position (then the point of the drop in the spectral weight at around the
quasiparticle dispersion kink). Obviously, there is a slope change in the real part of the antibonding total
self-energy ${\rm Re}\Sigma^{(2)}_{\rm tot}({\bf k},\omega)$. However, in the corresponding to this slope
change, a peak structure appears simultaneously in the imaginary part of the antibonding total self-energy
${\rm Im}\Sigma^{(2)}_{\rm tot}({\bf k},\omega)$. These results thus show that the quasiparticle dispersion
kink is induced by the slope change in ${\rm Re}\Sigma^{(2)}_{\rm tot}({\bf k},\omega)$, i.e., the position
of the quasiparticle dispersion kink shown in Fig. \ref{kink-maps} is exactly the same as that for the
corresponding inflection point in ${\rm Re}\Sigma^{(2)}_{\rm tot}({\bf k},\omega)$ shown in Fig.
\ref{band-structure}. This is why the kink in the quasiparticle dispersion marks the crossover between two
different slopes. Moreover, the position of the quasiparticle dispersion kink shown in Fig. \ref{kink-maps}
is also the exactly same as that for the corresponding peak in the imaginary part of the antibonding total
self-energy ${\rm Im}\Sigma^{(2)}_{\rm tot}({\bf k},\omega)$ shown in Fig. \ref{band-structure}, i.e.,
there is an exact one to one correspondence between the kink position shown in Fig. \ref{kink-maps} and the
peak position in ${\rm Im}\Sigma^{(2)}_{\rm tot}({\bf k},\omega)$ shown in Fig. \ref{band-structure}, which
therefore shows that the spectral weight at around the quasiparticle dispersion kink is reduced strongly by
the corresponding peak in ${\rm Im}\Sigma^{(2)}_{\rm tot}({\bf k},\omega)$, and then the weak spectral
intensity appears always at around the quasiparticle dispersion kink. On the other hand, the bilayer
coupling in Eq. (\ref{interlayer-hopping}) is absent at around the nodal region, and then the less visible
kink in the antibonding quasiparticle dispersion is caused mainly by the renormalization of the electrons
within a CuO$_{2}$ layer. However, at around the antinodal region, the bilayer coupling exhibits its largest
value, this large band splitting lead to that the break separating of the fast dispersive high-energy
part from the slow dispersive low-energy part more stronger. This strong separation of the quasiparticle
dispersion at the kink energy results in the enhancement of the kink effect in the quasiparticle dispersion
of the electron-doped bilayer cuprate superconductors at around the antinodal region. In other words, the
kink effect in the quasiparticle dispersion of the electron-doped bilayer cuprate superconductors at around
the antinodal region becomes the most pronounced due to the presence of the bilayer coupling.

In conclusion, within the framework of the kinetic-energy driven superconductivity, we have studied the
bilayer coupling effect on the quasiparticle dispersion kink in the electron-doped bilayer cuprate
superconductors. Our results show that the kink in the quasiparticle dispersion is present all around the
antibonding EFS, as the corresponding quasiparticle dispersion kink in the electron-doped single-layer
cuprate superconductors. However, in comparison with the corresponding case in the electron-doped
single-layer cuprate superconductors, the kink effect in the quasiparticle dispersion of the electron-doped
bilayer cuprate superconductors at around the antinodal region becomes the most pronounced, which therefore
indicates that the kink effect in the quasiparticle dispersion of the electron-doped bilayer cuprate
superconductors at around the antinodal region is enhanced by the bilayer coupling.

\section*{Acknowledgements}

ST is supported by Research Foundation of Yanshan University under Grant No. 8190448. HG is
supported by NSFC under Grant Nos. 11774019 and 12074022, and the Fundamental Research Funds for the
Central Universities and HPC resources at Beihang University. SF are supported by the National Key
Research and Development Program of China, and the National Natural Science Foundation of China (NSFC)
under Grant Nos. 11974051 and 11734002.

\end{document}